# A preliminary study of small-mass radiocarbon sample measurement at Xi'an-AMS[*]

Fu Yun-Chong(付云翀)[1,2]†, Zhou Wei-Jian(周卫健)[1,2,3], Du Hua(杜花)[1,2], Cheng Peng(程鹏)[1,2], Zhao Xiao-Lei(赵晓雷)[1,4], Liu Qi(刘起)[1,2], Lu Xue-Feng(卢雪峰)[1,2], Zhao Wen-nian(赵稳年)[2,3]

1) State Key Laboratory of Loess and Quaternary Geology, Institute of Earth Environment, Chinese Academy of Sciences, Xi'an, 710075

2) Shaanxi Key Laboratory of Accelerator Mass Spectrometry Technology and Application, Xi'an AMS center, Xi'an, 710061

3) Xi'an JiaoTong University, Xi'an 710049

4) A. E. Lalonde AMS Laboratory, University of Ottawa, 150 Louis Pasteur, Ottawa, ON., K1N 6N5, Canada

**Abstract**: To meet the measurement demands on small-mass radiocarbon (carbon content at $10^{-6}$g level) which are becoming increasingly significant. Xi'an-AMS has made improvement to the existing method of sample loading and has upgraded the Cs sputter ion source from the original SO-110 model. In order to study the feasibility of small-mass samples in Xi'an-AMS and evaluate the radiocarbon sample preparation ability using existing routine systems of $H_2$/Fe and Zn/Fe, the small-mass samples prepared by four different methods are tested. They are mass division method, mass dilution method, $H_2$/Fe reduction method and Zn/Fe reduction method. The results show that carbon mass above 25μg can be prepared using the existing Zn/Fe system, but no less than 100μg is required using the existing $H_2$/Fe system, which can be improved. This indicates Xi'an-AMS are now able to analyze small-mass radiocarbon samples.

**Keywords:** AMS，Small-mass radiocarbon sample，ion source

**PACS:** 07.75.+h, 29.25.-t, 29.90.+r

## I.    Introduction

Conventional decay counting and Accelerator Mass Spectrometry (AMS) are the two main analytical methods for radiocarbon[1]. Before the 1990's, decay counting was the dominant method of choice for the analysis of $^{14}$C, and the techniques had been developed to the point that the sample size needed for an analysis decreased from several grams to hundreds to dozens of milligrams (the gas sample method requires only 50mg; the liquid method, developed by Zhou Weijian, can handle sample as small as 100-200mg). The optimal precision of routine sample measurement can reach 2 ‰, and a single sample can be measured as quickly as within 1-2 days. Some highly efficient laboratories could analyze around 400 samples annually[2, 3]. Since the late 1970s, however, the small sample size needed for an analysis by the decay counting methods has been significantly surpassed with the emergence and subsequent rapid development of AMS technology. The carbon mass required for a routine analysis by AMS is only about 1mg, the time takes to complete a single-sample measurement can be less than one hour, the analysis precision of modern carbon can be better than 2 ‰, and over 1000 samples can be readily measured by an ordinary AMS-$^{14}$C laboratory[4, 5]. Therefore, it becomes essential to measure $^{14}$C samples using AMS; the decay counting method is now used much less frequently.

---
[*] Project supported by the National Natural Science Foundation of China under Grant No. 11205161.

† E-mail: fuyc@ieecas.cn

The hundreds to thousands fold reduction in sample size requirement by AMS has led to wide spread applications of $^{14}$C in numerous research fields such as earth science, environmental science, archeology, biosciences, marine science, cosmic chemistry and so on. Such expanding fields of research, that often require new and more difficult $^{14}$C analysis, continue to push for the need to measure ever smaller sample sizes. In many scientific studies, such as analyzing the radiocarbon in the air bubbles trapped in Antarctic ice, radiocarbon dating of rock paintings, micro-dose tracing in biosciences, "micro-damage" archeology, etc., one can only obtain very tiny amount of carbon ranging from a few to hundreds of micrograms (μg) [6-9]. Many AMS laboratories abroad, under the impetus of such demand, have started to focus on how to measure small-mass radiocarbon samples. One method is to use gas ion source to directly measure $CO_2$ gas samples by AMS. This requires substantial improvement on existing solid target ion sources or development of new gas ion sources with micro-flow controllable devices (μL/min level ) [10-13]. Another method is simply to push the limit of producing graphite with very small quantities of carbon, to be measured by existing solid target ion sources and AMS systems. Many laboratories have invested a great deal of effort in doing this. For example, ANSTO' scientists prepared 10-200μg samples by improved routine method and produced about 5μg samples by using 0.5mL laser-heated "microfurnace"; KCCAMS can produce and measure radiocarbon samples from approximately 0.015 to 0.1 mg; the Center for Applied Isotope Studies has developed a protocol for the measurement of radiocarbon in the range of 0.01-0.3mg[14-16]. Compared with the use of gas ion sources, making small-mass graphite is a relatively simpler approach, and it has several advantages such as stable beam intensity, higher efficiency, easy and quick sample change, low memory effect and so on [17] . There have been a number of AMS laboratories abroad undertaking small-mass radiocarbon research, but similar effort is currently still lacking in China, which has limited Chinese scientific research in certain areas.

This article will discuss the preliminary study on the feasibilities of carrying out routine small-mass radiocarbon samples measurement at Xi'an-AMS using the existing main equipment. We will present the result of a systematic assessment test, and identify the areas where further improvements are required.

## II.  Ion source and sample loading for small-mass radiocarbon sample measurement

The measurements of small-mass radiocarbon samples were carried out using the 3MV AMS system at the Xi'an-AMS Center, IEECAS. The key point for small-mass radiocarbon sample measurement is not only a suited ion source system, but also a powerful sample loading.

### 1)  Ion Source

Ion source system is often a key factor to determine whether small-mass radiocarbon samples can be measured. Compared to measuring more regularly sized samples, the measurement of small-mass samples could face some difficulties such as lower currents and elevated memory effect. Those laboratories pursuing in this direction often found it necessary to first carry out certain upgrade to their ion sources[15]. Xi'an-AMS uses a Cs sputter ion source of the SO-110 model from HVEE (High Voltage Engineering Europa), which is a major upgraded version from the manufacture's original design of 2008[18](See Figure 1a). The improved new ion source shown in Figure1(b) has its own distinctive character: the open structure around the ionizer and sample leads to better vacuum , an extra coolant loop is added to the ionizer supporting base, the halos surrounding the $Cs^+$ focusing spot is eliminated by shaping the edge of the ionizer central hole, the improved insulation structure provides full protection to the insulator. The new ion source's electrostatic and mechanical performance are more stable as a result of its improved pumping speed and carefully modeled temperature gradient throughout the ion source assembly. All these help to result in low cross contamination, which is essential for small-mass radiocarbon measurement.

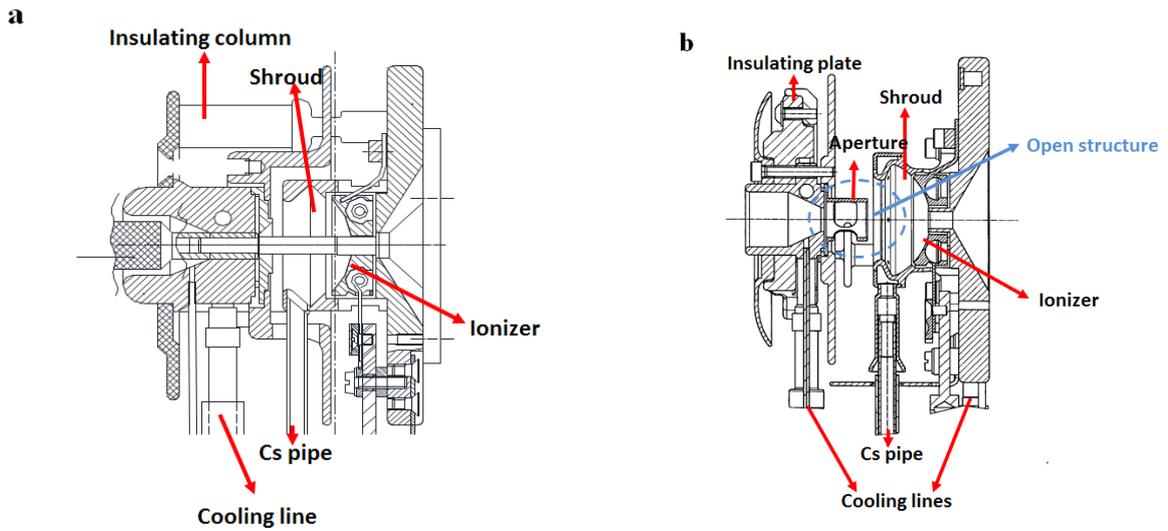

Before doing test measurement with small-mass radiocarbon samples, we first performed a SO-110 $Cs^+$ spot simulation using Simion8.0 and an actual $Cs^+$ measurement using a 3mm graphite rod target pressed into a 3mm diameter bore hole in a detachable aluminum target cap (Holder/Cathode). Figure 2(a) shows that $Cs^+$ beam has a poor focusing performance due to Simion's limitation of not including space charge calculation and lack of resolution on the ionizer's spherical surface, but from Figure 2(b), we can see that the $Cs^+$ spot actually has a diameter of 1.1mm (the black area within pale yellow dashed circle). The red dotted circle would be the regular sample size of 1.3mm-diamter. If a small-mass sample is packed into a 1mm-diamter hole, it should be fully sputtered, leading to higher utilization efficiency for the sample material. The spot center had a small offset, which can be corrected by careful ionizer assembly using a filler gauge.

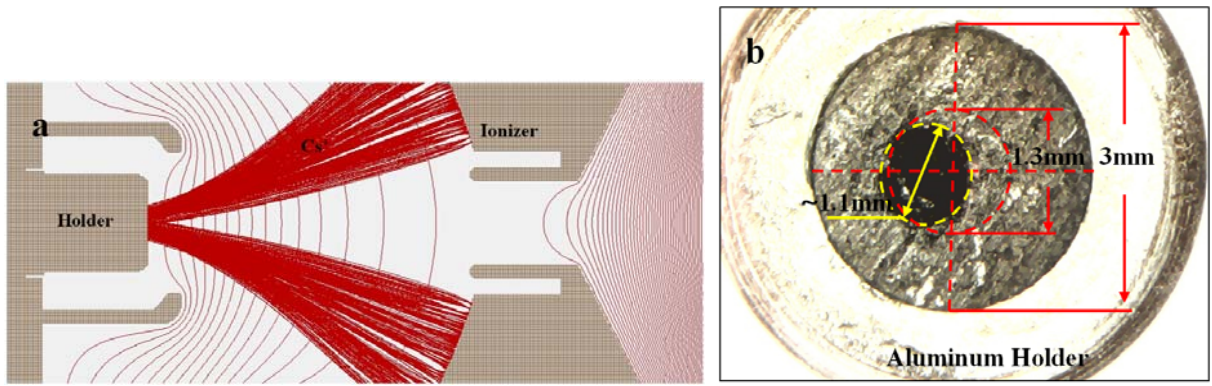

**2) Sample loading**

The mass and size of small-mass samples are less and far smaller than those of routine samples. This leads to difficulties in loading the sample into the holder [14]. The sample loading is completed using our designed and patented powder sample loading tool [19] (Figure3). Figure3 (a) shows the original sample loading and pressing tool, with which the powder sample is fed from the bottom of the fill, requiring many parts to be assembled to form a mold. After pressing one target, all parts of the device must be disassembled, cleaned, and reassembled again for the next target. The extrusion-needle is at the top of the press machine so that the direct impacts punch the extrusion-needle downward, thereby forming a sample mold. This process makes a direct impact on the long skinny extrusion-needle used, which is easily damaged. Also, when filling the powder, one couldn't observe the volume of the powder filled, easily causing mold rejection due to too much or too little sample in the holder. From Figure3 (b), we can see the redesigned sample loading and pressing tool which overcomes the disadvantages of original method by reversing the mold, that is, powder is filled from the top, using the backward extrusion to improve the efficiency of forming mold and to reduce wear of plunger chip and extrusion-needle and also minimize waste of sample material. The new method is particularly suitable for loading samples into small bore holes to simplify the handling of small-mass samples and to guarantee optimum sample density and surface finish for best ion source performance.

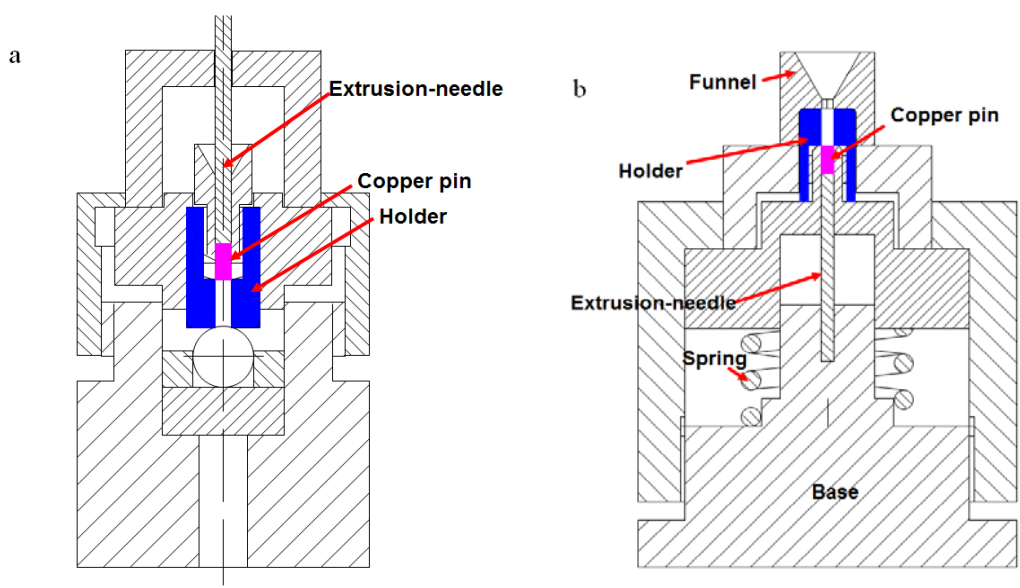

（color online）**Figure3 (a) Original tool of sample loading and pressing**      **(b) Redesigned tool of sample loading and pressing**

The redesigned sample loading device has made packing small-mass samples much easier to operate, but ultra small-mass sample loading is still extremely difficult to handle. Although one could consider increasing the proportion of the catalyst Fe powder, it is still very difficult to work with ultra small amount of $CO_2$. Therefore, we attempt to directly press Fe powder into copper holder first, which is then to be used directly in the reduction reaction, avoiding the difficulty of filling the power into a tiny hole after. Main reduction reaction of ultra small $CO_2$ gas is then performed at the holder surface by the catalyst Fe powder. Copper holder is chosen instead of routine aluminum holder for its higher melting point, because the holder is sealed in a quartz tube heated during the reduction process.

The improved SO-110 ion source, sample loading method and small bore hole holders were used in the measurements of small-mass samples in this work. Other operating parameters and the descriptions of Xi'an-AMS can be found in the reference [5].

# III. Preparation of small-mass radiocarbon samples[*]

The analysis of small-mass radiocarbon samples was done using the following 4 different methods.

**1）Mass division method.** Sample is prepared by the routine method to obtain conventional mg level graphite, which is then directly divided into small-mass samples of different mass (including 9 CSC samples from 20μg to 545μg). Different amount of Fe powder is filled into the bottom of a holder as the carbon mass is successively reduced. Because the graphitization is done with regular amount of $CO_2$, this method avoids possible preparation problems of small-mass radiocarbon, so it provides a relatively independent study on the measurement quality of small-mass radiocarbon samples using the Xi'an-AMS system. The problem with this method is that, it is hard to handle a few micrograms of extremely small-mass samples.

**2）Mass dilution method**. Sample is still prepared by the routine method to obtain conventional mg level graphite. Then the sample is divided into several parts and mixed with Fe powder in different proportion. The advantage of this method is that, it is easy to obtain ultra small-mass radiocarbon samples (including 5 CSC samples from about 2μg to 55μg and 8 blank samples from about 7μg to 531μg) and at the same time to avoid the complications in the graphitization process of small amount of $CO_2$. The beam intensity of this sample is close to the real situation of a directly reduced ultra small-mass sample, which mimics the performance of the Xi'an-AMS system when beam intensity is low. The simulated ultra small-mass sample is accumulated on the top surface of the holder with area smaller than $Cs^+$ spot, leading to the sample almost fully utilized, thus the real blank limitation of ultra small-mass samples can be derived.

3）**Direct preparation by the $H_2$/Fe reduction method.** The existing routine $H_2$/Fe reduction method and system are directly used to prepare small-mass radiocarbon (including CSC samples and background blank samples are 50μg, 100μg, 500μg, respectively). The advantage of the $H_2$/Fe method is its shorter reduction reaction time than the Zn/Fe method; the contamination possibility during preparation is minimized. This is the prime method used for small-mass radiocarbon preparation in the current study. The original samples (prior to division or dilution) used in the studies of method 1) and 2) are both prepared by the $H_2$/Fe method.

4）**Direct preparation by the Zn/Fe reduction method.** The existing routine Zn/Fe semi-automatic systems have 24 lines for reduction that have been established with assistance from University of Arizona. The Zn/Fe method is performed using Zn and $TiH_2$ as reductants in a few labs [15], but in our small-mass radiocarbon preparation the Zn quantity is adjusted to be 30mg and the Fe/C ratio is far more than the regular 2/1. Various samples under 100μg were prepared using the existing Zn/Fe system, including 4 OXⅡ samples from 50μg to 68μg and 3 same known-samples of the FIRI code 'C' (the Fourth International Radiocarbon Inter-Comparison) from 25μg to 50μg .

Fe powder is also a key factor for the measurement of small-mass radiocarbon sample; therefore we experimented with different Fe powders so that we can choose the best for the purpose. In addition, we prepared a special type of small-mass radiocarbon samples by means of Fe powder direct pressed in copper holders for their direct use in reduction reactions. All details will be discussed in the next section.

# IV. Result and Discussion

Four kinds of small-mass samples were measured at Xi'an-AMS[*]. Figure 4 shows the pMC (percent Modern Carbon) values (a), the average $^{12}C^{3+}$ beam intensity (b), $^{13}C/^{12}C$ (c) and $(^{14}C/^{12}C)/(^{13}C/^{12}C)^2$ (d) of CSC standards by different preparation methods using $H_2/Fe$ system with mass from few to hundreds micrograms. From Figure 4, the pMC of 'Mass division method' of the CSC standards are relatively stable, which only slightly decline with mass decreased, and the $^{12}C^{3+}$ intensity obviously decreases with mass but still stays above 8μA at least. Sample sizes are separated into smaller mass range using 'Mass dilution method'. The pMC are slightly wavy at beginning and consistent with pMC values of 'Mass division method' above 20μg range, then acutely decrease with mass under 20μg; the $^{12}C^{3+}$ intensity almost has the same tendency. This demonstrates that the pMC and beam intensity suffer severely decline with the decrease of sample size, especially when it is approaching 20μg. Minami *et al* explain this as caused by varying isotopic fractionations due to differences in graphitization yield and target thickness [20], but in these measurements described here, the original samples were obtained by routine preparation methods with typical mg mass, which were just divided or diluted by extra Fe powder mechanically. Figure 4 (b) indicates that the beam intensity is obviously decreasing along with sample size; this phenomenon is particularly evident at ultra small-mass under extreme low beam conditions. This implies that pMC has a certain correlation with beam intensity, which will be studied in next experiments. From Figure 4, we can see the results of direct preparation by the $H_2/Fe$ reduction method of sample mass of 500μg, 100μg and 50μg. The pMC of 500μg and 100μg samples are consistent with preceding methods, and $^{12}C^{3+}$ beam intensities are also declining with mass. Then, the 50μg sample has huge deviation from the normal value. This sample gave no $^{12}C^{3+}$ beam (only equivalent to charge digitizer noise) and few $^{14}C$ counts (few counts which can be expected from pure Fe powders) during AMS measurement; therefore, the so-called 50μg sample actually had no carbon in it. It is a result of low graphitization yield and lack of pressure measurement sensitivity in chemical preparation process when $CO_2$ pressure runs too low. From Figure4(c) and (d), we can see that the trend of $^{13}C/^{12}C$ ratio is a mirror image of pMC and the ratio of $(^{14}C/^{12}C)/(^{13}C/^{12}C)^2$ also have the same trend as pMC. However, $^{13}C/^{12}C$ ratios of 100 and 500μg directly prepared samples are slightly lower than division samples of similar mass, and 50μg sample failed also in Fig.4(c) and (d). Consequently, the sample less than 100μg is not doable using the existing routine $H_2$/Fe system; we need to improve this system in the next step. The experimental results indicate that the small-mass samples should be calibrated to standards of the same carbon content level. In future analysis, small-mass samples and standards for normalization will be grouped by different mass ranges, to be analyzed in the same batch together. The sample mass can be roughly divided into ranges of <20μg, 20—80μg, 80—300μg and >300μg.

---

[*] Note: The pMC of division and dilution methods samples were calibrated by regular standard as unknown samples. The $^{12}C^{3+}$ beam intensity of regular analysis is typically between 30-40μA at Xi'an-AMS.

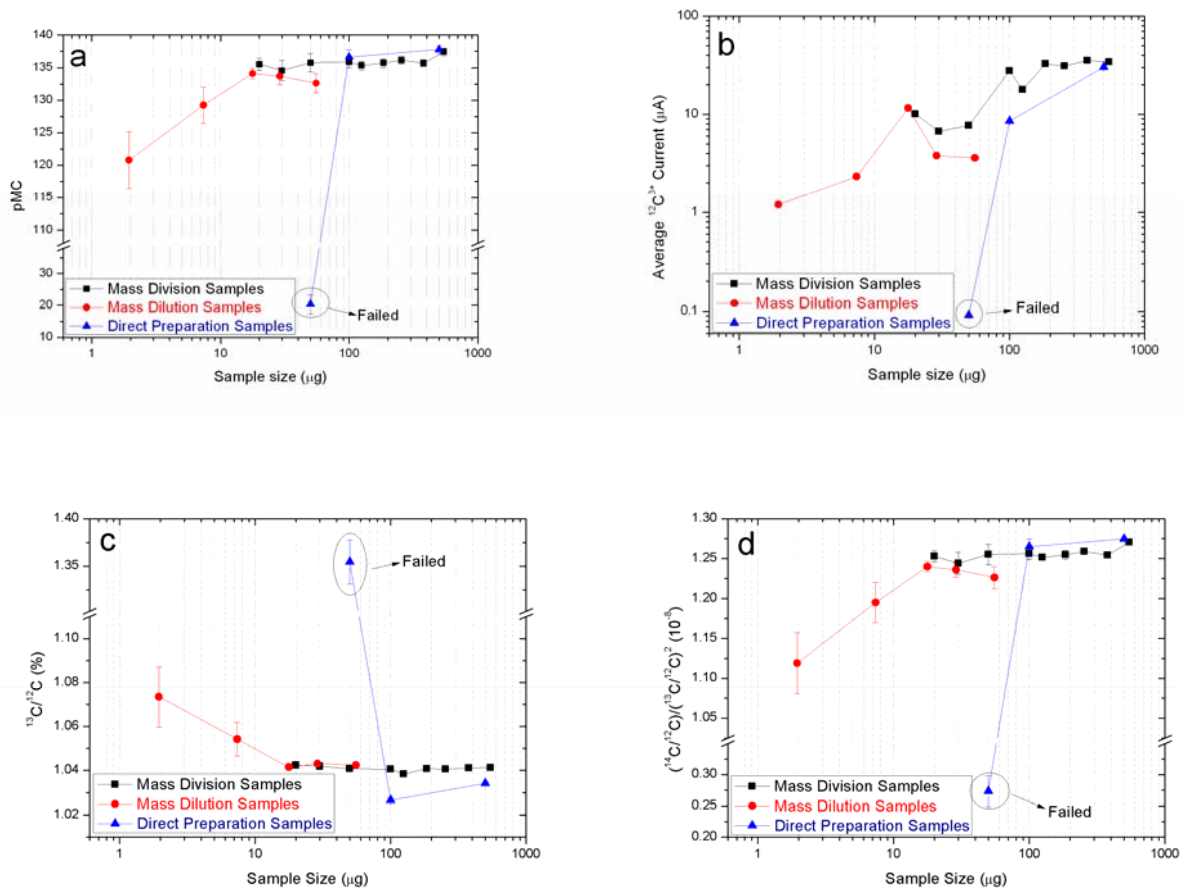

（color online）**Figure 4 The pMC values (a), the average beam intensity (b), $^{13}C/^{12}C$(c) and $(^{14}C/^{12}C)/(^{13}C/^{12}C)^2$ (d) of CSC standards by different methods versus sample size**

The blank samples were obtained by two methods. Figure 5 shows the pMC (a)，the average beam intensity (b)，$^{13}C/^{12}C$(c) and $(^{14}C/^{12}C)/(^{13}C/^{12}C)^2$ (d) of Blind Coal blank samples versus sample size. As mentioned above, the limitation to the ultra small-mass sample measurement blank can be assessed using 'Mass dilution method' samples, and the real situation for existing chemical preparation system can be reflected by the directly prepared samples. Fig.5 (a) shows the results for sample sizes from about 7μg to 531μg (compared to CRC samples, the blanks made by the dilution method can span a larger mass range from a few micrograms to half milligrams). With the reduction of sample size, the pMC increases gradually. We can regard it as the way to assess the size limitation to the small-mass samples (It is deducible that the actual directly prepared ultra small-mass blank samples will be worse than what can be accounted for just from the measurement using the Xi'an-AMS system). The minimum blank sample size that can be $^{14}$C dated is 7.5μg, to correspond to 34607 ± 703a BP. Figure 5 also shows the results of direct preparation by the $H_2$/Fe reduction method of samples 500μg, 100μg and 50μg, and a 900μg regular blank sample. Obviously, the pMC results of directly prepared samples are slightly higher than same mass sizes by the dilution method samples, but they have the same trend. Figure5 (c) indicates that two directly prepared samples of ≤500μg are slightly lower than dilution samples of similar mass. We can learn from figures 4(C) and 5(C) that isotopic fractionation was caused by small-mass chemical preparation processes in existing $H_2$/Fe system. From Figure5 (d), we can see that the $(^{14}C/^{12}C)/(^{13}C/^{12}C)^2$ also has the same trend as pMC. The 50μg blank sample of direct preparation by the $H_2$/Fe reduction method also failed.

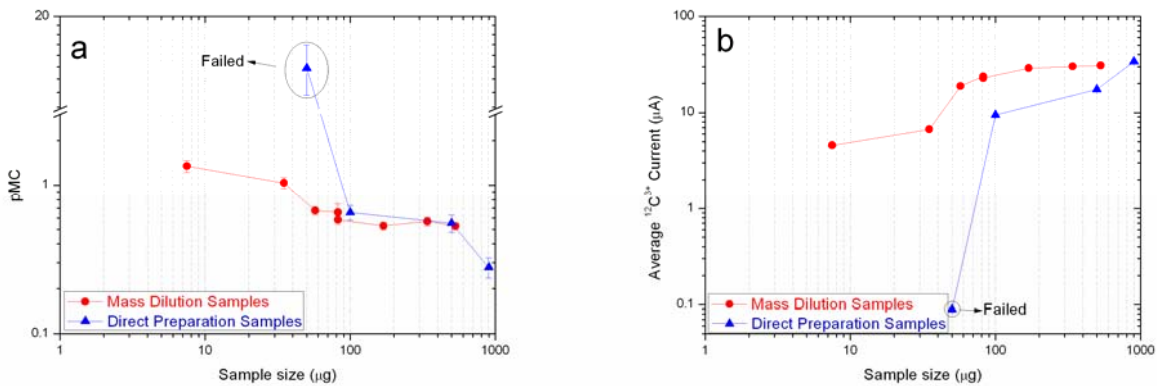

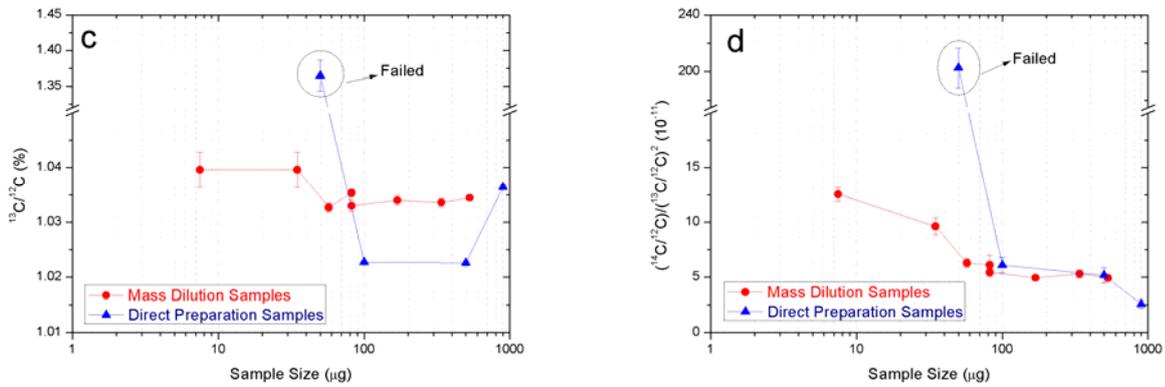

（color online）**Figure 5 The pMC values (a), the average beam intensity (b), $^{13}C/^{12}C$(c) and $(^{14}C/^{12}C)/(^{13}C/^{12}C)^2$ (d) of Blind Coal blanks by different methods versus sample size**

Figure 6 shows the pMC of small-mass OXⅡ standards and known samples prepared by the Zn/Fe system using small-mass standards calibration with sample size. Only samples under 100μg are discussed here. The known samples were prepared from the preserved sample of the FIRI code 'C' sample. The population mean of this known sample is 18132±245a and AMS mean is 18175±135a[21]. From Figure 6, we can see that the pMC of OXⅡ standards are stable and the known samples are consistent within 1σ, but the error relatively increases as the measurement time being just 300 seconds, far less than normal measurement time. Table 1 shows the results of standards and known samples using different calibration methods. From Table 1, our results of small-mass samples and the reported value are consistent within the error. Therefore, the existing Zn/Fe system has the ability to prepare small-mass sample above 25μg.

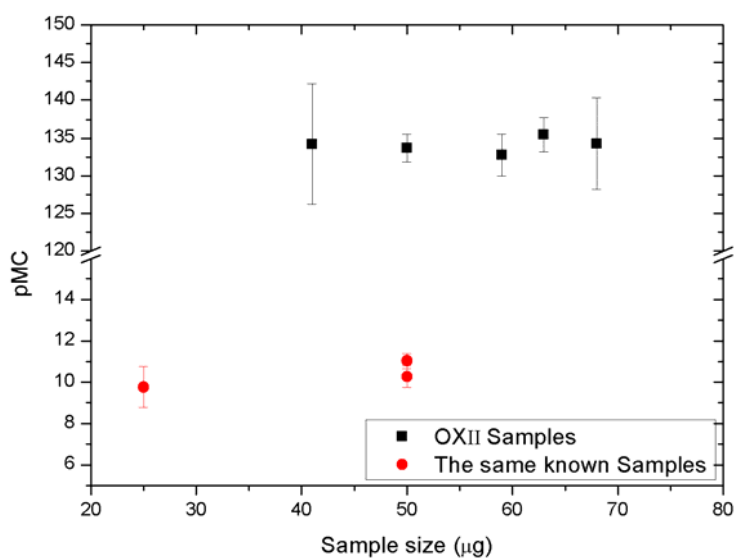

（color online）Figure 6 The pMC values of small-mass OXⅡ standards and known samples prepared by Zn/Fe system using small-mass

standards calibration versus sample size

Table.1  The AMS results of small-mass standard and known samples preparation by Zn/Fe system using

different calibration method[*1]

| Target No. | Description of Sample | Mass (μg) | Regular size standard calibration | | | | Small-mass standard calibration | | | |
| | | | pMC | | [14]C Age（a BP） | | **pMC** | | **[14]C Age (a BP)** | |
| | | | pMC | Error (1σ) | [14]C Age | Error (1σ) | **pMC** | **Error (1σ)** | **[14]C Age** | **Error (1σ)** |
| XAT0107 | OXⅡ | 63 | 133.03 | 2.33 | - | - | **135.45** | **2.28** | **-** | **-** |
| XAT0108 | OXⅡ | 59 | 130.38 | 2.81 | - | - | **132.75** | **2.79** | **-** | **-** |
| XAT0109 | OXⅡ | 41 | 131.78 | 7.80 | - | - | **134.17** | **7.97** | **-** | **-** |
| XAT0110 | OXⅡ | 50 | 131.32 | 1.93 | - | - | **133.71** | **1.85** | **-** | **-** |
| XAT0111 | OXⅡ | 68 | 131.87 | 5.93 | - | - | **134.26** | **6.04** | **-** | **-** |
| XAT0116 | Known Sample | 25 | 9.59 | 0.98 | 18832 | 780 | **9.76** | **1.00** | **18690** | **781** |
| XAT0117 | Known Sample | 50 | 10.84 | 0.36 | 17849 | 263 | **11.03** | **0.36** | **17707** | **261** |
| XAT0118 | Known Sample | 50 | 10.09 | 0.52 | 18427 | 407 | **10.27** | **0.53** | **18284** | **406** |

| XAT0116-0118 Mean Value | 10.17 | 0.25 | 18360 | 196 | **10.35** | **0.25** | **18218** | **192** |
|---|---|---|---|---|---|---|---|---|

In addition to the above four kinds of small-mass sample preparation methods studied, effects of different Fe powders were also studied. The impact of Fe powder can almost be ignored for measurement of regular mg sized samples, but with the decrease of sample size, the influence of Fe powder increases gradually. Therefore, choice the right Fe powder catalyst is crucial to small-mass samples measurement. We experimented with routinely used domestic Fe powder of 99% purity and the Sigma-Aldrich 99.995% $Fe_2O_3$ (SA $Fe_2O_3$). Those powders were reduced by $H_2$ or direct pressed into holder. 22 $^{14}C$ counts can be detected in 300 seconds using routine Fe powder reduced by $H_2$, approximately 0.073 atoms/sec; 9 $^{14}C$ counts in 300s can be detected from SA $Fe_2O_3$ reduced by $H_2$, approximately 0.030 atoms/sec; the sample of directly pressed SA $Fe_2O_3$ in a holder gives no $^{14}C$ count in 300s (Note: routine modern radiocarbon standard sample gives $^{14}C$ counting rate of about 100 atoms/sec). Obviously, SA $Fe_2O_3$ is better than currently used domestic Fe powder. In the next step, we will study more catalyst material so that a suitable catalyst for small-mass samples measurement can be found for both high purity and low cost. The $^{14}C$ counting rate from SA $Fe_2O_3$ reduced by $H_2$ is higher than that from direct pressed SA $Fe_2O_3$ due to vacuum limitations of the $H_2/Fe$ system or $H_2$ purity. This means that the ultra small-mass samples are not doable using the existing $H_2/Fe$ system.

Furthermore, two samples of Fe powder were direct pressed in copper holders for direct use in reduction reactions. The $^{12}C^{3+}$ beam intensity of XAT0120 (500μg) is slightly more than $1/9^{th}$ that of a regular massed sample, and the same is true for XAT0121 (300μg). These results show that the beam intensities of this method are smaller than those of similar sized samples prepared by other methods. The main reason is that reaction is limited to the limited pressed target surface area only. After reduction reactions were completed we observed the black powder adhering to quartz tube wall, which was considered to be the 'excess' carbon not able to combine with Fe powder. Probably for ultra small-mass samples carbon could completely combine with Fe powder, because in this case the carbon content is far smaller than the reduction capacity of the pre-pressed Fe surface. The results show that $^{14}C/^{12}C$ is about $1.379\times10^{-12}$, $^{13}C/^{12}C$ is about $1.0287\times10^{-2}$, $^{12}C^{3+}$ is about 3.8μA, which are similar to the regular sized CSC samples ($^{14}C/^{12}C$ is about $1.417\times10^{-12}$,

$^{13}C/^{12}C$ is about $1.0397\times10^{-2}$, $^{12}C^{3+}$ is about 32μA). All these samples were measured similarly in a 200s quick preliminary test batch. Therefore, this method is feasible. However, we need to study the impact on $^{14}C$ background from the copper holders entering the reduction reactions and to further improve $H_2/Fe$ system for ultra small-mass samples.

## V.    Conclusion and Future Plans

According to the experimental results and the analysis the following conclusions can be drawn:

1) Xi'an-AMS totally has the ability to measure small-mass radiocarbon sample, after it has improved sample loading method and upgraded the ion source that provided feasibility foundation for the analysis. In order to utilize the small-mass samples more efficiently, the 1mm-diameter holder will be used. In future, the $^{12}C$ and $^{13}C$ dark current influence on measurement will be further investigated.

2) The method of Fe powder being directly pressed in copper holders for direct use in reduction reactions, is suited for handling ultra small-mass samples and to avoid loading, but the effect on $^{14}C$ background must be evaluated and the procedures optimized in the next step.

3) Small-mass radiocarbon samples are best calibrated with standards at closer carbon content levels for AMS measurement.

4) Small-mass radiocarbon sample of mass less than 100μg can not be prepared using the existing $H_2/Fe$ system. In future, we will build a new one or improve the existing system. Basically the reduction-unit need to be redesigned as follows: Firstly, improve vacuum of the system; secondly, increase pressure sensor sensitivity, and reduce volume of the reaction tube; furthermore, choose more pure reagents and so on to achieve ultra/small-mass radiocarbon samples preparation.

5) Small-mass radiocarbon samples above 25μg can be prepared by the existing Zn/Fe system, which can satisfy the preparation of majority of the small-mass samples. However, this system needs to be improved further for ultra small-mass samples preparation. The reduction tube volume needs to be reduced by half and the pressure sensor needs to be replaced with a more accurate and more sensitive one. It can be expected that ultimate measurable small carbon mass can break 10μg.